\documentclass[aps, prb, showpacs, preprint, amsmath, letterpaper]{revtex4-1}

\usepackage{graphics}
\usepackage{graphicx}
\usepackage{amssymb}
\usepackage{bm}

\begin{document}
\title{Expansion of the high field-boosted superconductivity in UTe$_{2}$ under pressure}

\author{Sheng Ran$^{1,2,3}$, Shanta R. Saha$^{1,2}$, I-Lin Liu$^{1,2}$, David Graf$^{4}$, Johnpierre Paglione$^{1,2}$, Nicholas P. Butch$^{1,2*}$}
\affiliation{$^1$ Maryland Quantum Materials Center, Department of Physics, University of Maryland, College Park, MD 20742, USA
\\$^2$ NIST Center for Neutron Research, National Institute of Standards and Technology, Gaithersburg, MD 20899, USA
\\$^3$ Department of Physics, Washington University in St. Louis, St. Louis, MO 63130, USA
\\$^4$ National High Magnetic Field Laboratory, Florida State University, Tallahassee, FL 32313, USA
\\$*$ email: nbutch@umd.edu
}

\date{\today}

\begin{abstract}
Magnetic field induced superconductivity is a fascinating quantum phenomenon, whose origin is yet to be fully understood. The recently discovered spin triplet superconductor, UTe$_{2}$, exhibits two such superconducting phases, with the second one reentering in the magnetic field of 45~T and persisting up to 65~T. More surprisingly, in order to induce this superconducting phase, the magnetic field has to be applied in a special angle range, not along any high symmetry crystalline direction. Here we investigated the evolution of this high-field induced superconducting phase under pressure. Two superconducting phases merges together under pressure, and the zero resistance persists up to 45~T, the field limit of the current study. We also reveal that the high field-induced superconducting phase is completely decoupled from the first order field polarized phase transition, different from previously known example of field induced superconductivity in URhGe, indicating a superconductivity boosted by a different paring mechanism.

\end{abstract}
\maketitle

\section*{Introduction}
The recent evidence for spin triplet superconductivity in UTe$_{2}$ has opened an avenue for the study of topological superconductivity~\cite{Ran2019}. The superconducting state of UTe$_{2}$ closely resembles that of ferromagnetic superconductors, but the normal state is paramagnetic and shows no indication of magnetic ordering~\cite{Sundar2019,Hutanu2020,Duan2020}. Spin triplet pairing is strongly indicated by the extremely large, anisotropic upper critical field $H_{\textup{c2}}$~\cite{Ran2019,Aoki2019}, nodes on the superconducting gap~\cite{Metz2019,Bae2021}, and the temperature independent NMR Knight shift in the superconducting state~\cite{Nakamine2020,Nakamine2021}. The superconducting order parameter comprises two components and breaks time reversal symmetry~\cite{Hayes,Agterberg2020}. A nontrivial topology is suggested by the observation of chiral in-gap bound states by scanning tunneling spectroscopy~\cite{Jiao2020}. 

Even more striking, UTe$_{2}$ hosts two independent field-induced superconducting phases~\cite{Ran2019a,Knebel2019,Knafo2020,Knafo2021}, with the higher-field phase reentering at a high magnetic field of 45~T and persisting up to 65~T when the magnetic field is aligned over a limited angular range about the normal direction of the (011) plane. The quasi-two-dimensional Fermi surface revealed by band structure calculations and photoemission measurements~\cite{Miao2020,Xu2019,Ishizuka2019}, as well as the lack of a ferromagnetic ground state, has led to suggestions that the field-induced superconductivity in UTe$_{2}$ is due to reduced dimensionality instead of magnetic fluctuations~\cite{Lebed1998,Mineev2020}: a magnetic field applied parallel to quasi 2D conducting layers will stabilize superconductivity when the magnetic energy reaches the hopping amplitude between the conducting layers~\cite{Lebed1998,Lebed2021}. 

In this work we investigate the evolution of the magnetic field-induced superconducting phases in UTe$_{2}$ as pressure is applied to samples oriented specifically along the off-axis angle which stabilizes the high-field phase. Over a range of pressures near 1~GPa, the two different superconducting phases merge together, and the electrical resistance remains zero up to at least 45~T, a remarkably large value for a superconductor with a 3~K critical temperature. The high field-induced superconducting phase is completely decoupled from the first-order transition to a field-polarized state, suggesting that magnetic fluctuations may not be crucial to this reentrant superconductivity. At pressures exceeding the critical pressure at which metamagnetic phase transition extrapolates to zero magnetic field, we observe features with the same temperature dependence as the high field-induced superconducting phase, further investigation of which might shed light on the mechanism reentrant superconductivity.

\section*{Results and discussion}
\subsection*{Pressure-magnetic field phase diagram}

\begin{figure}
\includegraphics[angle=0,width=160mm]{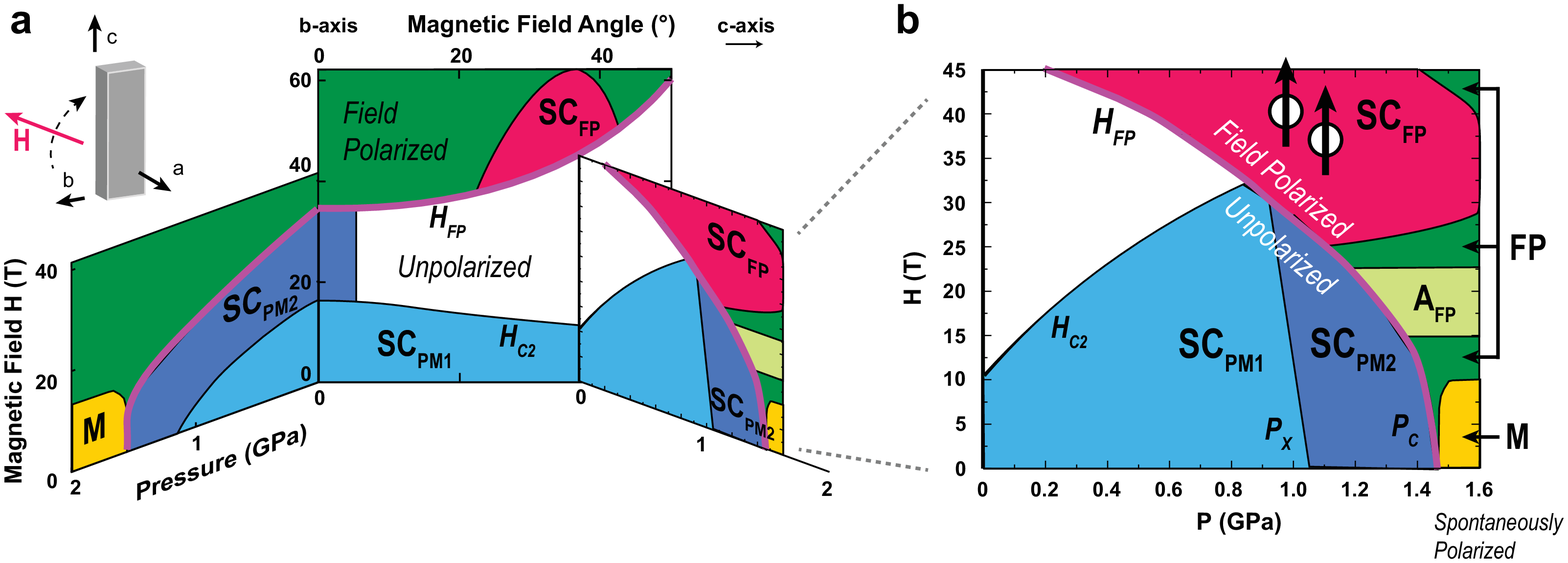} 
\caption{\textbf{The stability of superconducting and magnetic phases of UTe$_2$ as a function of magnetic field strength and angle, and applied pressure.} a) The zero-field superconducting phase SC$_{PM1}$ exhibits a strongly direction dependent upper critical field $H_{C2}$. Along the crystallographic $b$-axis, the reentrant superconducting phase SC$_{PM2}$ is stabilized up to a first-order magnetic transition at $H_{FP}$, above which exists a magnetically polarized phase FP. Applied pressure uniformly suppresses all field scales and $H_{FP}$ bounds superconductivity up to a critical pressure $P_c$. b) For a magnetic field oriented between 25 and 30 degrees between the $b$- and $c$-axes, the high-field reentrant superconductivity SC$_{FP}$ stabilizes in the FP phase. Applied pressure enhances SC$_{PM1}$, which meets a decreasing $H_{FP}$ at a crossover region $P_x$. Here, SC$_{PM1}$ transitions to SC$_{PM2}$, which survives up to $P_c$, where it is replaced by magnetic order M. Below $P_x$, $H_{FP}$ is a lower bound on SC$_{FP}$, but above $P_x$, the two are decoupled and SC$_{FP}$ survives beyond P$_c$. An anomaly A$_{FP}$, suggestive of Landau-level superconductivity, emerges above $P_c$. The phase diagram is based on the resistance data shown in Fig.~2.} 
\label{Fig1}
\end{figure}

To characterize the evolution of the high field-induced superconducting phase under pressure, we performed complementary measurements of electrical resistance $R$ and tunnel diode oscillator (TDO) frequency $\Delta f$, which is sensitive to the change of both electrical resistance and magnetic susceptibility. Two samples were studied for which the magnetic field was applied approximately 25 degrees and 30 degrees, respectively, away from the $b$ axis towards $c$. At ambient pressure, three distinct phases were observed as shown in Fig.~\ref{Fig1}: a field polarized state FP with greatly enhanced magnetization and resistance~\cite{Ran2019a,Miyake2020,Knafo2020,Knafo2021}; the low field superconducting phase, SC$_{\textup{PM}}$, coexisting with the paramagnetic state; and the high field-induced superconducting phase, SC$_{\textup{FP}}$, existing inside the field polarized state. Criterion used to infer critical magnetic field for each phase are explained in the Supplementary Information (See Supplementary Figure 4-7).  

Absent in this field configuration is another field induced superconducting phase that is observed on the low-field side of the metamagnetic field $H_{\textup{FP}}$ when magnetic field is applied along the $b$ axis. For magnetic field along the $b$ axis, applied pressure suppresses the metamagnetic field to zero at 1.5~GPa and forces a phase transition inside the paramagnetic state at a crossover pressure $P_\textup{x}$, from SC$_{\textup{PM}1}$ to SC$_{\textup{PM}2}$, in zero magnetic field at approximately 1~GPa~\cite{Lin2020}. Recent symmetry analysis indicates that the SC$_{\textup{PM}1}$ phase has two-component order parameter while order parameter of SC$_{\textup{PM}2}$ phase only has one component~\cite{Agterberg2020}. Our measurements in this study do not exhibit any field-induced features~\cite{Lin2020}, so we infer that the SC$_{\textup{PM}1}$-SC$_{\textup{PM}2}$ boundary is nearly vertical in the $H-P$ plane. At pressures exceeding 1.5~GPa, long range magnetic order is stabilized~\cite{Ran2020,Braithwaite2019,Thomas2020,Aoki2021}, whose features have been interpreted in terms of both ferromagnetism~\cite{Ran2020, Lin2020} and antiferromagnetism~\cite{Thomas2020}. Without an unambiguous experimental proof of the nature of this phase, in this study we label it as a magnetically ordered phase, M. 

\begin{figure}
\includegraphics[angle=0,width=160mm]{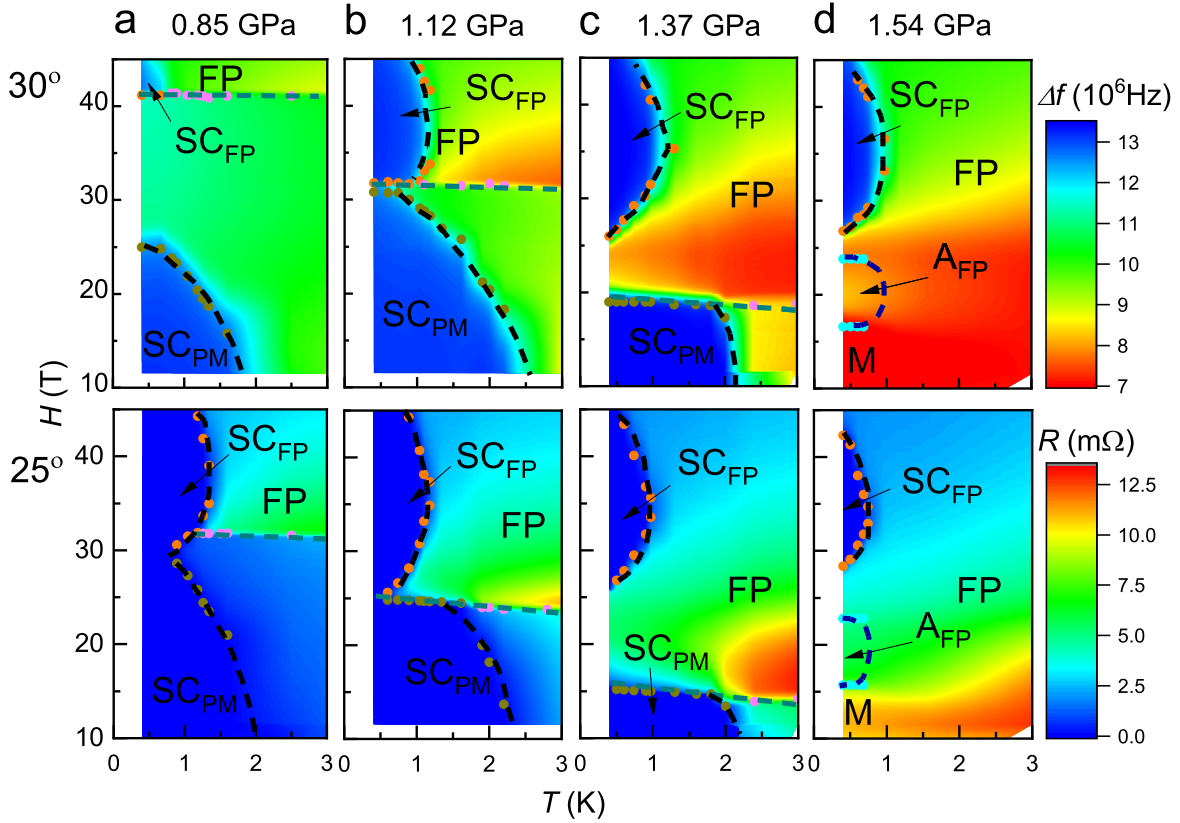}
\caption{\textbf{High-field tunnel diode oscillator (TDO) frequency (top row) and resistance (bottom row) at different pressures.} TDO frequency is sensitive to the change of both electrical resistance and magnetic susceptibility of the sample. Magnetic field is applied at 30 degree from $b$ axis towards $c$ axis for TDO measurement, and 25 degree from $b$ axis towards $c$ axis for resistance measurement. a) In the low-pressure region, $H_{FP}$ serves as a lower bound to SC$_{FP}$, whose dome is cut off discontinuously. b) At crossover pressures $P_x$, $H_{FP}$ falls approximately between SC$_{FP}$ and SC$_{PM}$, whose ranges of stability would otherwise overlap. c) Above $P_x$, $H_{FP}$ serves as the upper field limit for SC$_{PM}$. d) Above $P_c$, long-range ordered magnetism M sets in while SC$_{FP}$ survives and the anomalous feature A$_{FP}$ emerges as a local maximum in $\Delta f$. Solid dots are from experimental data, dashed lines are guide for the eye. Note that TDO and resistance measurements are performed for slightly different magnetic field direction, which leads to the slight different phase diagrams for each pressure.}
\label{Fig2}
\end{figure}

The pressure-magnetic field phase diagram for the magnetic field in this angle range is summarized in Fig.~\ref{Fig1}b. All three phases manifest clear evolution under pressure, as seen in $R$ and $\Delta f$ in Fig.~\ref{Fig2}. The metamagnetic field is monotonically suppressed by the applied pressure, similar to the behavior for field along $b$~\cite{Lin2020}, although it starts at a higher value. The metamagnetic field vanishes at a critical pressure $P_\textup{c}$ between 1.47 and 1.54~GPa, giving rise to a spontaneous polarized state in zero magnetic field beyond this pressure. Over the entire pressure range, both $R$ and $\Delta f$ change discontinuously on passing through the metamagnetic field in the normal state, implying that it maintains the first-order metamagnetic transition observed at ambient pressure~\cite{Miyake2020,Knafo2020,Imajo2020}.  

\begin{figure}
\includegraphics[angle=0,width=160mm]{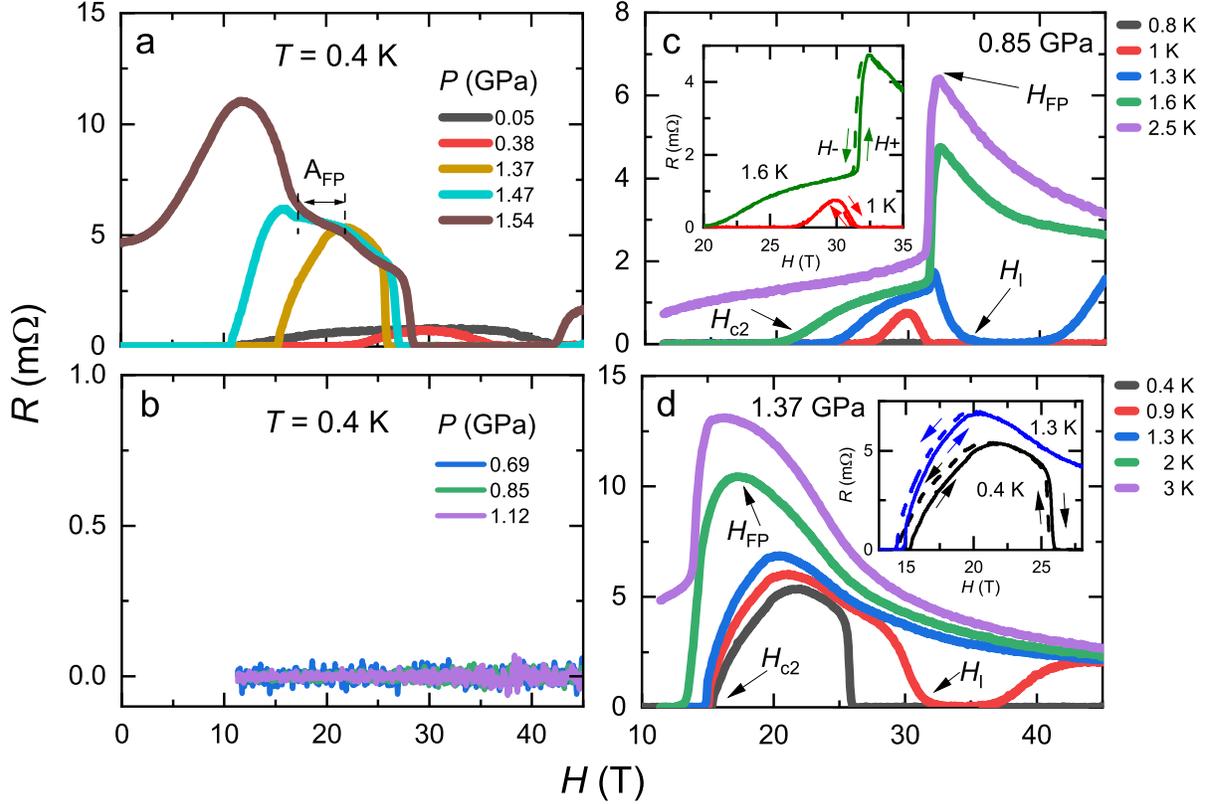} 
\caption{\textbf{High-field magnetoresistance, highlighting different phase boundaries at different pressures.} Magnetic field is applied at 25 degree from $b$ axis towards $c$ axis. a) Pressure-dependence of 0.4 K magnetoresistance, showing the sharpening of phase boundaries at higher pressures and emergence of high-pressure phase A$_{FP}$. b) The resistance at pressures 0.69, 0.85, and 1.12 GPa is zero between 11~T and 45~T at 0.4 K. Zero resistance persists to zero magnetic field at these pressure values. c) For fields lower than $H_{FP}$, the superconducting transitions $H_{C2}$ are broad, but $H_{FP}$ is sharp. As shown in inset, $H_{FP}$ is hysteretic, reflecting the 1st order transition, both at high temperatures in the FP phase and low temperature in the SC$_{FP}$ phase. d) Much sharper superconducting phase boundaries occur once $H_{FP}$ limits SC$_{FP}$. As shown in inset, low-field phase boundaries are hysteretic and first order, as is the onset of SC$_{FP}$.}
\label{Fig3}
\end{figure}

Upon initial increase of the pressure, the stability of the both superconducting phases is enhanced: the upper critical field of SC$_{\textup{PM}}$, $H_{\textup{c}2}$, increases and the critical onset field of SC$_{\textup{FP}}$, $H_\textup{l}$, which coincides with the metamagnetic field, decreases. In an intermediate crossover pressure range, the phase boundary between SC$_{\textup{PM}}$ and SC$_{\textup{FP}}$ is no longer visible in the electrical resistance; this remains zero up to 45~T at base temperature, which is noteworthy as it is the largest DC magnetic field currently available to experiment (Fig.~\ref{Fig3}b). As the pressure further increases, the upper critical field of SC$_{\textup{PM}}$ is limited by the metamagnetic field and decreases, but the critical onset field of SC$_{\textup{FP}}$ starts to increase, and the two superconducting phases are no longer connected. When the metamagnetic field vanishes, SC$_{\textup{PM}}$ is suppressed completely. 

Subtle differences between the electrical resistance and TDO samples highlight the effects of their slight angular offset (Fig.~\ref{Fig2}). As the TDO sample sits at a slightly larger angle away from the $b$ axis, the upper critical field of SC$_{\textup{PM}}$ has a smaller value and critical onset field of SC$_{\textup{FP}}$ has a larger value at ambient pressure. The SC$_{\textup{PM}}$ and SC$_{\textup{FP}}$ phases remain connected over a much smaller pressure range. Similarly, the field polarized state starts above a higher magnetic field because the metamagnetic field is larger, but it is suppressed faster upon increasing pressure, and vanishes also at $P_\textup{c}$.  

The discontinuous nature of the metamagnetic transition at $H_{\textup{FP}}$ is conspicuous in the resistance data. At low pressures (Fig.~\ref{Fig3}c), a sharp upward jump marking the FP phase boundary is replaced at low temperatures by a sharp downward jump marking the SC$_{\textup{FP}}$ phase boundary. An additional hallmark of first-order transitions, namely field-hysteresis is also readily apparent (inset of Fig.~\ref{Fig3}c). In the intermediate pressure range, $H_{\textup{FP}}$ limits the lower-field superconducting phase SC$_{\textup{PM}}$ (Fig.~\ref{Fig3}d). Here, the transitions associated with $H_{\textup{FP}}$ are again hysteretic (inset of Fig.~\ref{Fig3}d). Interestingly, the decoupled SC$_{\textup{FP}}$ phase appears to also exhibit some small hysteresis, the origin of which may be associated with details of the field-reentrance. These features are consistent at all measured pressures (Fig.~\ref{Fig3}a, b).

These details reveal important points about the relationships between the many electronic phases. The low- and high-field superconducting phases always exist on opposite sides of the metamagnetic transition $H_{\textup{FP}}$, upon which Fermi surface reconstruction has been suggested based on thermoelectric power and Hall effect measurements~\cite{Miyake2020,Niu2020,Niu2020a} . In addition, in the low pressure range, SC$_{\textup{FP}}$ only appears on the high-field side of the metamagnetic field (Fig.~\ref{Fig2}a, b), while SC$_{\textup{PM}}$ only exists at fields lower than the metamagnetic field. The role of the metamagnetic field switches in the intermediate pressure range, where the metamagnetic field truncates the low-field phase SC$_{\textup{PM}}$. This behavior is apparent in both panels of Fig.~\ref{Fig2}c, where the high-temperature part of the upper critical field of SC$_{\textup{PM}}$ curve rises rapidly as temperature decreases, extrapolating to field values far higher than the metamagnetic field, but then, as soon as the upper critical field of SC$_{\textup{PM}}$ coincides with the metamagnetic field, the behavior suddenly changes and the upper critical field of SC$_{\textup{PM}}$ becomes almost temperature independent down to low temperatures. Taken together, these facts imply that the SC$_{\textup{PM}}$ and SC$_{\textup{FP}}$ phases separated by the metamagnetic transition might have different superconducting pairing that are unique to PM and FP, respectively. It is particularly striking that both low- and high-field superconducting phases look like they would cover much larger ranges of field were they not limited by the metamagnetic field.

In the high pressure range, the critical onset field of SC$_{\textup{FP}}$ and the metamagnetic field are well separated, by more than 20~T. This is crucial for the understanding of the pairing mechanism of the SC$_{\textup{FP}}$ phase. In the case of URhGe, the reentrance of superconductivity can be explained in terms of ferromagnetic fluctuations parallel to the direction of the magnetic field. In that case, reentrant SC occurs in the vicinity of the magnetic critical field. In UTe$_{2}$ at low pressure, the SC$_{\textup{FP}}$ phase resembles somewhat the reentrant phase in URhGe, leading to the speculation that magnetic fluctuations are also responsible for reentrant superconductivity in UTe$_{2}$. However, here we show clearly that the SC$_{\textup{FP}}$ phases can exist in the region far away from the field polarized phase line, indicating a possibility scenario that magnetic fluctuations are not responsible for the pairing mechanism. Future experiments to investigate magnetic fluctuations in the vicinity of the SC$_{\textup{FP}}$ phase at pressure above 1.3~GPa will potential shed light on the pairing mechanism. 

\subsection*{Anomaly in the high-pressure region}
\begin{figure}
\includegraphics[angle=0,width=160mm]{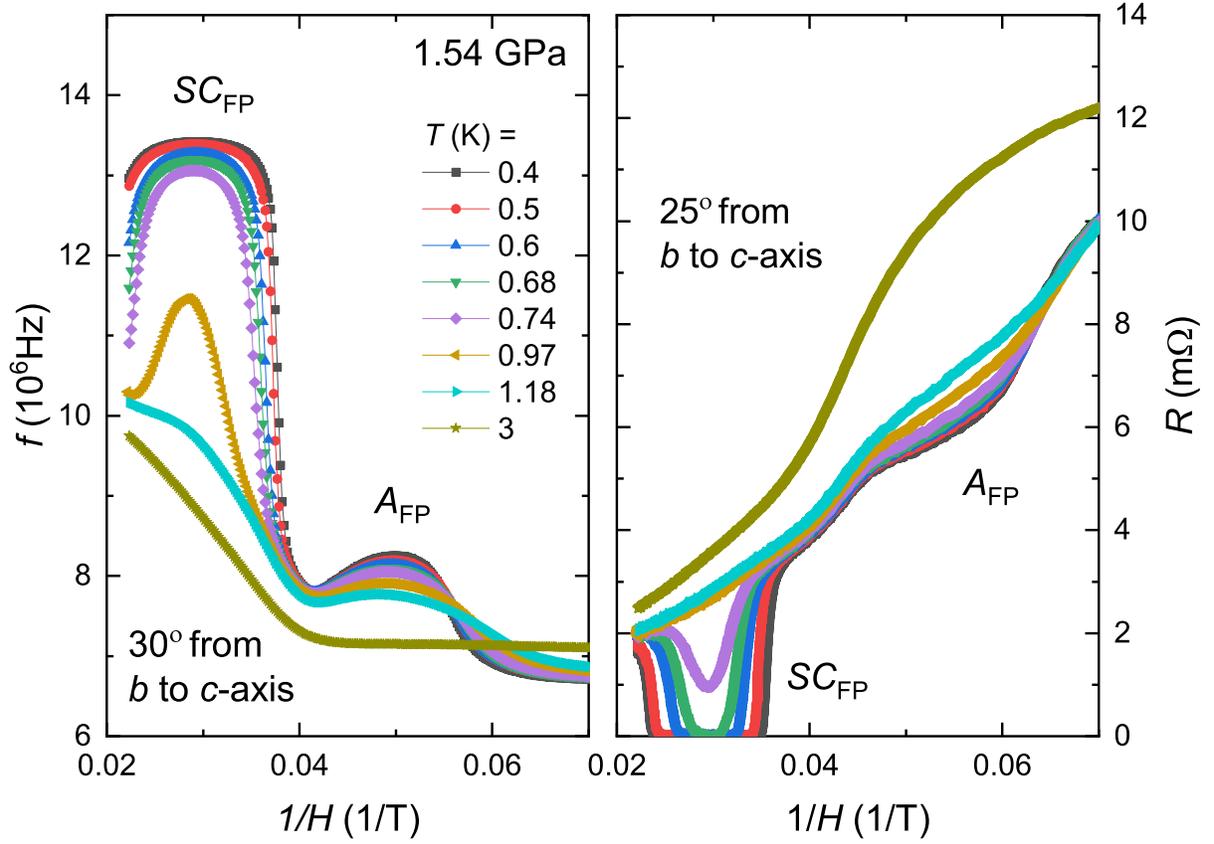} 
\caption{\textbf{Additional anomalies emerge in the high pressure region}. A$_{FP}$ phase appears at 1.54 GPa in both TDO and resistance data, which have the same temperature dependence as $SC_{FP}$ phase, indicating A$_{FP}$ phase might be the precursor of superconductivity. Magnetic field is applied at 30 degree from $b$ axis towards $c$ axis for TDO measurement, and 25 degree from $b$ axis towards $c$ axis for resistance measurement.}
\label{Fig4}
\end{figure}

A striking characteristic of the high-pressure FP phase is the emergence of additional features in the field range between the M and SC$_{\textup{FP}}$ phases. These anomalies, denoted A$_{\textup{FP}}$, are pronounced in the $\Delta f$ data, and noticeable in $R$ data (Fig.~\ref{Fig4}). It is not clear whether these anomalies correspond to a thermodynamically distinct phase. In order to trace the evolution of these anomalies, we introduced a criteria to define the boundaries as shown in the Supplementary Materials. These anomalies exhibit a clear temperature dependence below 1.2~K. This notable similarity to the temperature dependence of SC$_{\textup{FP}}$ suggests that A$_{\textup{FP}}$ is a closely related phenomenon, with a similar energy scale to that of the superconductivity. This distinguishes A$_{\textup{FP}}$ from the zero-field magnetically ordered phase M, which has a higher-temperature ordering temperature relative to superconductivity, that continues to increase with applied pressure.

An exciting possibility is that A$_{\textup{FP}}$ is a precursor to superconductivity. Previous theoretical studies have shown that in extreme magnetic field Landau levels will have dramatic influence on the low-temperature behavior of the upper critical field~\cite{Gruenberg1968,Tesanovic1989,Song2017}. Indeed, a more recent theoretical study indicates that SC$_{\textup{FP}}$ might be a form of superconductivity that is enhanced by high-field Landau quantization of the conduction electrons~\cite{Park2020}. Therefore, A$_{\textup{FP}}$ and SC$_{\textup{FP}}$ may actually be the same superconducting phase occurring at different Landau levels, analogous to Shubnikov-de Haas oscillations. A challenge to this interpretation is that A$_{\textup{FP}}$ is not accompanied by a zero resistance state (Fig.~\ref{Fig3}a), but a plausible explanation for this is that A$_{\textup{FP}}$ is actually partially superconducting due to the effects of energy-level broadening are stronger at lower field. Based on the inverse-field periodicity of A$_{\textup{FP}}$, the next Landau level will be centered at approximately 100~T, achieving zero resistance at magnetic fields as low as 90~T, which is practically achievable using the strongest available non-destructive pulsed magnet systems.

\section*{Methods}
\subsection*{Crystal synthesis}
Single crystals of UTe$_{2}$ were synthesized by the chemical vapor transport method using iodine as the transport agent. Elements of U and Te with atomic ratio 2:3 were sealed in an evacuated quartz tube, together with 3~mg/cm$^3$ iodine. The ampoule was gradually heated up and hold in the temperature gradient of 1060/1000~$^{\circ}$C for 7 days, after which it was furnace cooled to room temperature. The crystal structure was determined by $x$-ray powder diffraction using a Rigaku $x$-ray diffractometer with Cu-K$_{\alpha}$ radiation. Crystal orientation was determined by Laue $x$-ray diffraction performed with a Photonic Science $x$-ray measurement system. 
\subsection*{Measurement}
Magnetoresistance and tunnel diode oscillator (TDO) measurements were performed at the National High Magnetic Field Laboratory, Tallahassee, using the 45-T hybrid magnet. A non-magnetic piston-cylinder pressure cell was used for measurements under pressure up to 1.57 GPa, with Daphne oil 7575 as the pressure medium. Pressure was calibrated at low temperatures by measuring the fluorescence wavelength of ruby, which has a known temperature and pressure dependence. The TDO technique uses an $LC$ oscillator circuit biased by a tunnel diode whose resonant frequency is determined by the values of $LC$ components, with the inductance $L$ given by a coil that contains the sample under study; the change of its electrical resistance and magnetic properties results in a change in resonant frequency. Identification of commercial equipment does not imply recommendation or endorsement by NIST. Error bars correspond to an uncertainty of one standard deviation.

\section*{DATA AVAILABILITY}
The data that support the findings of this study are available from the corresponding author upon reasonable request.

\section*{Acknowledgments}
We acknowledge D. Agterberg, G. S. Boebinger, A. E. Koshelev, A. Lebed, and V. P. Mineev for inspiring discussions. Research at the University of Maryland was supported by the National Institute of Standards and Technology (NIST), the US National Science Foundation (NSF) Division of Materials Research Award No. DMR-1610349, the US Department of Energy (DOE) Award No. DE-SC-0019154 (experimental investigations), and the Gordon and Betty Moore Foundation’s EPiQS Initiative through Grant No. GBMF4419 (materials synthesis). Work performed at the National High Magnetic Field Laboratory, USA (NHMFL) was supported by NSF through NSF/DMR-1644779 and the State of Florida. 

\section*{Author contributions}
N. Butch and S. Ran conceived and designed the study. S. Ran and S. R. Saha synthesized the single crystalline samples. S. Ran, S. R. Saha, I. Liu and D. Graf performed the electrical resistivity and tunnel diode oscillator measurements. S.R., J.P. and N.P.B. analyzed the data and wrote the paper with everyone's contribution.

\section*{Competing Interests}
The authors declare no competing interests. 

\bibliography{UTe2HF}

\bibliographystyle{naturemag}

\clearpage

\end{document}